**Complexity science approach to economic crime**

János Kertész[1] and Johannes Wachs[2,3]

[1]Department of Network and Data Science, Central European University, Quellenstrasse 51, A-1100 Vienna
[2]Complexity Science Hub Vienna, Josefstädter Strasse 39, A-1080 Vienna
[3]Institute for Information Business, Vienna University of Economics and Business, A-1020 Vienna

**Standfirst:** János Kertész and Johannes Wachs discuss how complexity science and network science are particularly useful for identifying and describing the hidden traces of economic misbehaviour such as fraud and corruption.

The rules society makes for the economy and its agents are often broken. Corruption, collusion, fraud, tax evasion, and other forms of misbehaviour are widespread and costly. The UN estimates[a] that the world-wide annual cost of corruption is over $US3 trillion. These economic losses are compounded by downstream effects including increasing inequality and decreasing social trust. Certainly, corruption has received plenty of attention from researchers and policymakers alike. Yet it is unclear if these collective efforts are helping. A recent review by Global Financial Integrity suggests that, according to Transparency International's Corruption Perceptions Index and the World Bank's Control of Corruption Index, corruption levels are no better than they were 20 years ago in over two-thirds of countries.

Help may come from complexity science. In the past two to three decades a growing body of research has demonstrated the potential of complexity and network science approaches to the study of various economic phenomena (Box 1). Because corruption and economic crimes are themselves complex social and economic processes involving many actors and agents, it is becoming clear that such approaches are highly relevant to the tasks of detecting, describing, and controlling corruption.

The fundamental obstacle to a better understanding of corruption remains measurement. Indeed, corruption is supposed to be kept secret by the perpetrators and so can only be tracked indirectly, in a wider sea of data. Data on company ownership or on links between managers and politicians are not readily available. Many existing data sources, unlike corrupt actors, respect national borders and are not easily compared internationally. Any study of corruption using proven or ground-truth cases must contend with selection bias. Traditional studies comparing countries rely on surveys, asking citizens or businesspeople about their perceptions of and experiences with corruption, an approach known to suffer from biases and limitations of scale[1].

Fortunately, the situation is improving. Governments and international organizations are making big administrative data available. Enterprising researchers, investigative journalists, and NGOs are crawling massive amounts of social and economic data on firms and their owners. They are in some way mapping slices of the whole economy in their search for traces or patterns of corrupt behaviour. Recent research carried out on a variety of these datasets has perhaps one unifying strand: the distribution and organization of such suspicious behaviour is highly complex and far from random.

---

[a] https://news.un.org/en/story/2018/12/1027971

An important example in this genre is the study of public procurement markets. On these markets governments purchase goods and services from the private sector, amounting to 10–20% of GDP in OECD countries. Data about public procurements are openly accessible and contain much information about the specificities of the bids, enabling systematic quantitative analysis of corruption risk with the tools of network science.

Procurement procedures are typically highly regulated but complicated enough that skilled corrupt officials can steer contracts to insider firms, avoiding competition and ensuring generous margins. In fact, the share of contracts awarded without competition in a country correlates significantly with traditional survey-based corruption measures[2]. Mapping procurement markets as networks of contract issuers and winning firms reveals that these high corruption risk contracts are significantly clustered: if one suspicious deal is uncovered, another is likely nearby.

Firms themselves have intricate relationships between each other, and corporations and individuals use sophisticated structures of ownership, parent firms and subsidiaries to minimize their tax bills. As highlighted by the so-called Panama Papers[b] such arrangements are often legal grey areas. Therefore, it is vital to map out the global ownership relations between firms. Recent work applied network science methods to such data to highlight the jurisdictions that serve as so-called sinks and conduits of the tax-offshoring industry[3]. Though a great deal of money ends up in the sinks, which are the stereotypical island tax havens, it turns out that the conduits through which money flows on the way to the sinks are often larger countries, including the Netherlands and the UK. A recent and controversial preprint published by the World Bank discovered a link between aid payments to developing countries and financial in-flows into tax havens[4]. The study applied traditional econometric models to convincingly relate the timing of these flows. How such money gets from the World Bank to a bank account on the Cayman Islands is likely a complex story.

In some cases, the analysis is more focused, dealing with a single country, for example. Thorough investigation of traditional news media and Wikipedia material has been used to reveal the evolving network of Brazilian politicians involved in corruption scandals[5]. Over the 27 years of observation, the network of corrupt actors evolves — most actors are involved in only one scandal, but a few key actors bridge scandals across governments and parties. Such a longitudinal, network-based approach is able to identify the key players in the emergence of a giant component of crooks. A systematic study of a major Mexican scandal using tools of network science has shown that the dynamic aspect is indeed important for the identification of metrics sensitive to corruption risk[6].

Identifying warning signals of misbehaviour in economic activity is one of the main goals of this type of research, and we expect a major impact on the fight against corruption. For example, network science provides tools to select companies suspicious of collusion from public procurement data[7]. As data quality and linkages between datasets improve, methods of complexity science are likely to provide further useful insights. This is because economic crimes tend to leave faint but interrelated traces across the different datasets in which they appear.

---

[b] https://www.icij.org/investigations/panama-papers/

Despite the great potential of these new datasets and methods, complexity scientists researching corruption need to engage with sociology if they want to move beyond descriptive studies of corruption to predicting it and understanding the underlying mechanisms. Sociologists understand that the complexity of corruption is rooted in social processes and that hypotheses about corruption should start there. Multi-disciplinary cooperation is needed to test and further develop models based on those hypotheses and to validate them with the help of empirical data.  Agent-based models are a particularly useful tool to explore such processes and represent a method that social scientists have already used to understand macro phenomena such as norms and social conventions that cannot easily be explained by micro behavior[8]. The stubborn persistence of corruption in some regions is one such macro outcome that surely merits further study from this perspective. Network scientists and physicists have a great deal to contribute to this effort at the nexus of data, modelling, and theory.

**Acknowledgement:** JK acknowledges support from the Hungarian Scientific Research Fund (OTKA K129124).

**References:**
1. Heywood, P. M. & Rose, J. Close but no Cigar: the measurement of corruption. *J. of Pub. Policy*. **34**, 3 (2014).
2. Wachs, J., Fazekas, M. & Kertész, J. Corruption risk in contracting markets: a network science perspective. *Int. J. Data Sci. Anal.* (2020).
3. Garcia-Bernardo, J., Fichtner, J., Takes, F.W., & Heemskerk, E.M. Uncovering Offshore Financial Centers: Conduits and Sinks in the Global Corporate Ownership Network. *Sci. Rep.* **7**, 6246 (2017).
4. Andersen, J.A., Johannesen, N., & Rijkers, B. Elite Capture of Foreign Aid, Evidence from Offshore Bank Accounts. *World Bank Policy Research Working Paper*, **9150** (2020).
5. Ribeiro, H.V., Alves, L.G.A, Martins, A.F., Lenzi, E.K., & Perc, M. The dynamical structure of political corruption networks. *J. Comp. Netw.* **6**, 6 (2018).
6. Luna-Pla, I. & Nicolás-Carlock, J.R. Corruption and complexity: a scientific framework for the analysis of corruption networks. *Appl. Netw. Sci.* **5**, 13 (2020).
7. Wachs, J. & Kertész, J. A network approach to cartel detection in public auction markets. *Sci. Rep.* **9**, 10818 (2019).
8. Macy, M. W., & Willer, R. From factors to actors: Computational sociology and agent-based modeling. Annual Review of Sociology, 28, 1 (2002).
9. Arthur, W.B. Complexity and the economy. *Science*. **284**, 107-109 (1999).
10. Battiston, S., Puliga, M., Kaushik, R., Tasca, P., & Caldarelli, G. Debtrank: Too central to fail? Financial networks, the Fed and systemic risk. *Sci. Rep.* **2**, 541 (2012).

**Box 1: Complexity in economics**
The economy is a paradigmatic example of a complex system, with its many agents from individuals and households to corporations, banks and the state itself, connected to each other by a diversity of interactions[9].  If complex behaviour emerges from simple statistical physics models in which the "agents" interact locally and obey simple rules, it is no wonder that more extreme forms of complexity are found in the economy where individual behaviour is so rich. Yet mainstream economics abstracted away these idiosyncrasies by considering wise representative agents. Physicists recognize this as a "mean field" approximation of the system. Such approaches work surprisingly well in average situations but can fail catastrophically as seen during the 2008 financial crisis.

In recent decades, and especially since the global crisis, researchers have increasingly turned to the tools of complexity science to understand the economy. An example is DebtRank[10], a new networked measure of systemic risk in financial markets, inspired by the algorithms that power internet search engines.

One must be careful in making analogies between the economy and physical complex systems. "Microscopic laws" for economic behaviour are far from being fully explored. Indeed, the rich field of behavioural economics demonstrates that actors significantly deviate from models of rational behaviour. Economic agents strategize and often consider the effects of their actions. On top of this, an "uncertainty principle" is at play: observations on the economy influence its evolution, for example when an arbitrage opportunity is discovered.